\pdfoutput=1 
\documentclass[
superscriptaddress,
nofootinbib,
 aps,
reprint
]{revtex4-1}

\usepackage{graphicx}
\usepackage{dcolumn}
\usepackage{bm}
\usepackage{amsmath}
\usepackage{color}
\usepackage{float}
\usepackage{array,url,kantlipsum}
\usepackage{verbatim}
\usepackage{xcolor}
\usepackage{mathtools, cuted}
\usepackage[mathlines]{lineno}

\newcolumntype{P}[1]{>{\centering\arraybackslash}p{#1}}

\begin{document}
\preprint{APS/123-QED}

\title{Electrical readout microwave-free sensing with diamond}
\author{Huijie Zheng$^\dagger$}
\email{zheng@uni-mainz.de}
\affiliation{Johannes Gutenberg-Universit{\"a}t Mainz, 55128 Mainz, Germany}
 \affiliation{Helmholtz-Institut, GSI Helmholtzzentrum f{\"u}r Schwerionenforschung, 55128 Mainz, Germany}
 
\author{Jaroslav Hruby}
\thanks{These authors contributed equally to this work.}
\affiliation{IMOMEC division, IMEC, Wetenschapspark 1, B-3590 Diepenbeek, Belgium}
\affiliation{Institute for Materials Research (IMO), Hasselt University, Wetenschapspark 1, B-3590 Diepenbeek, Belgium.}

\author{Emilie Bourgeois}
\affiliation{IMOMEC division, IMEC, Wetenschapspark 1, B-3590 Diepenbeek, Belgium}
\affiliation{Institute for Materials Research (IMO), Hasselt University, Wetenschapspark 1, B-3590 Diepenbeek, Belgium.}

\author{Josef Soucek}
\affiliation{IMOMEC division, IMEC, Wetenschapspark 1, B-3590 Diepenbeek, Belgium}
\affiliation{Czech Technical University in Prague, Sitna sq. 3105, 272 01, Kladno, Czech}

\author{Petr Siyushev}
\affiliation{Institute for Quantum Optics and IQST, Ulm University, Albert-Einstein-Allee 11, D-89081 Ulm, Germany}

\author{Fedor Jelezko}
\affiliation{Institute for Quantum Optics and IQST, Ulm University, Albert-Einstein-Allee 11, D-89081 Ulm, Germany}

\author{Arne Wickenbrock}
\affiliation{Johannes Gutenberg-Universit{\"a}t Mainz, 55128 Mainz, Germany}
 \affiliation{Helmholtz-Institut, GSI Helmholtzzentrum f{\"u}r Schwerionenforschung, 55128 Mainz, Germany}
 
\author{Milos Nesladek}
\affiliation{IMOMEC division, IMEC, Wetenschapspark 1, B-3590 Diepenbeek, Belgium}
\affiliation{Institute for Materials Research (IMO), Hasselt University, Wetenschapspark 1, B-3590 Diepenbeek, Belgium.}
\affiliation{Czech Technical University in Prague, Sitna sq. 3105, 272 01, Kladno, Czech}

\author{Dmitry Budker}
\affiliation{Johannes Gutenberg-Universit{\"a}t Mainz, 55128 Mainz, Germany}
 \affiliation{Helmholtz-Institut, GSI Helmholtzzentrum f{\"u}r Schwerionenforschung, 55128 Mainz, Germany}
\affiliation{Department of Physics, University of California, Berkeley, California 94720, USA}

 \date{\today}

\begin{abstract}
While nitrogen-vacancy (NV$^-$) centers have been extensively investigated in the context of spin-based quantum technologies, the spin-state readout is conventionally performed optically, which may limit miniaturization and scalability. Here, we report photoelectric readout of ground-state cross-relaxation features, which serves as a method for measuring electron spin resonance spectra of nanoscale electronic environments and also for microwave-free sensing. As a proof of concept, by systematically tuning NV centers into resonance with the target electronic system, we extracted the spectra for the P1 electronic spin bath in diamond. Such detection may enable probing optically inactive defects and the dynamics of local spin environment. 
We also demonstrate a magnetometer based on photoelectric detection of the ground-state level anticrossings (GSLAC), which exhibits a favorable detection efficiency as well as magnetic sensitivity. 
This approach may offer potential solutions for determining spin densities and characterizing local environment.


\end{abstract}

\pacs{Valid PACS appear here}
\maketitle


Nitrogen-vacancy (NV) centers in diamond have become a prominent platform for spin-based quantum technology, being utilized to study small-scale quantum information processing concepts and quantum sensing under ambient conditions\,\cite{barry2020sensitivity,reiserer2016robust}. Conventionally, the readout of the spin state is based on the optically detected magnetic resonance (ODMR) technique and NV center visualization is performed optically, taking advantage of the high-yield photoluminescence of the NVs. This typically requires setups built on optical tables to achieve the best performance. In that sense, fabrication of miniaturized compact devices and their integration for practical uses is a complex engineering endeavor\,\cite{siyushev2019photoelectrical,doherty2016towards}. Alternatively, electrical readout is a convenient way to measure the spin state of a qubit, and recently, it has been successfully applied to NV centers via, for example, photoelectric detection of magnetic resonance (PDMR)\,\cite{bourgeois2015photoelectric}, as well as photoelectric readout of electron\,\cite{hrubesch2017efficient} and nuclear\,\cite{morishita2020room,gulka2017pulsed} spins. With this detection method, NV$^-$ centers can be addressed with a spatial resolution limited by the nanoscale feature size of electron-beam lithography\,\cite{siyushev2019photoelectrical,bourgeois2015photoelectric,manfrinato2013resolution,hrubesch2017efficient,gulka2017pulsed}, which could enable spin readout in fabricated dense arrays, and therefore, providing fully integrated quantum-diamond solutions. The advantage over optical readout is that photocurrent (PC) detection allows one to overcome the spatial resolution limitation imposed by the diffraction limit. Moreover, electrical readout outperforms optical means in collection efficiency, as the latter are limited by finite-efficiency objective optics and the high index of refraction of diamond, and it is typically on the order of a percent of lower\,\cite{siyushev2019photoelectrical}. Photoelectric detection can achieve high efficiency since all the carriers can be collected and multiplied due to photoelectric gain\,\cite{bourgeois2015photoelectric}.
However, an important problem is the microwave (MW) induced noise and cross-talks\,\cite{gulka2017pulsed} which limit the sensitivity of magnetic field sensing and imaging and also provides engineering and technological hurdles for a high-level integration.

Here, we propose and implement a novel microwave-free PDMR detection for monitoring the local NV electron spin dynamics within the diamond lattice; in this scenario, we study spin transitions at the ground-state level anticrossings (GSLAC), occurring at the fields of $\approx$ 102.4\,mT, which lead to narrow resonances and can be employed for sensitive magnetometry. The role of nuclear spins in these transitions and pinpoint differences to microwave-free detection is discussed in Ref.\,\cite{gulka2021room}. The developed methodology also opens perspectives for highly integrated magnetometry with electrical readout.

We begin with the investigation of magnetic resonances that originate from ground-state level anticrossing (GSLAC) and cross relaxation by performing both photoelectric and optical readout. 
Based on these results, we designed and implemented a magnetometer based on the PC detection of the GSLAC. This magnetometer alleviates the requirement of microwave control\,\cite{zheng2019microwave} and PC readout enables better magnetic sensitivity per volume compared to optical detection. 



\begin{figure}
\centering
\includegraphics[width=\columnwidth]{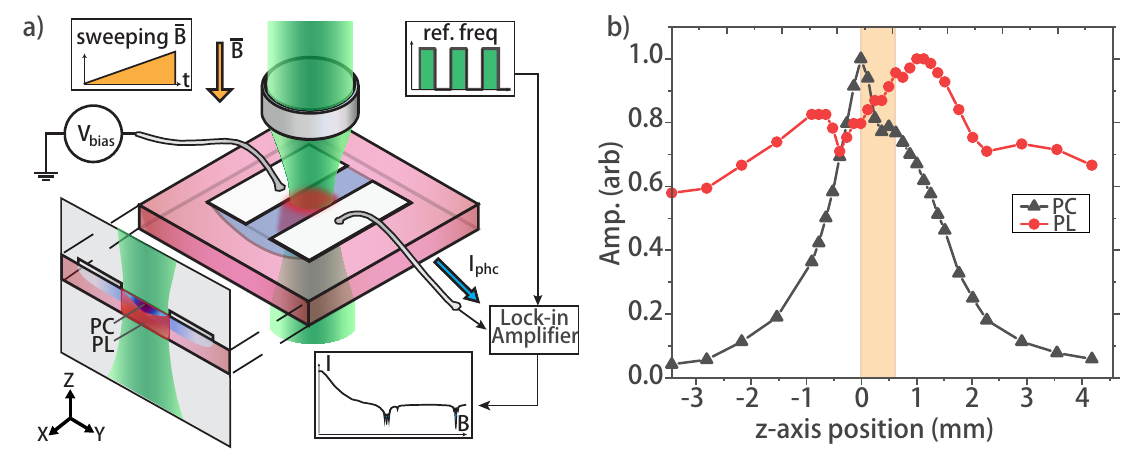}
\caption{(a) Schematic representation of the setup used for PC detection. A static magnetic field ($B_s$) is applied along an NV axis (the $z$ direction). A green laser beam is focused in between a pair of 5\,$\mu$m separated lithographic electrodes. The PC signal, $I_{phc}$, is detected with a lock-in amplifier with reference to the modulation frequency of the laser light by an acousto-optic modulator (AOM). A projected side view indicates the effective detection volume of PC (blue region) and PL (red region) readout.
(b) The signals of PC (black solid triangles) and PL (red solid circles) readout when the focus lens is moved along the light-propagation direction ($z$-axis, perpendicular to the diamond surface). The diamond sample is indicated with an amber rectangle. 
}
\label{fig:schematic}
\end{figure}

\textit{Photoelectric detection of NV$^-$ magnetic spectra}.--
Throughout the experiments a high-pressure high-temperature (HPHT) diamond was used (see Appendix B). To collect charge carriers under bias voltage, coplanar Ti/Al electrodes with a 5\,$\mu$m gap were lithographically patterned on the diamond surface. The photocurrent was pre-amplified and recorded via lock-in amplifier (LIA) with a reference to the laser modulation frequency. A custom-made electromagnet was built to provide the background magnetic field \cite{zheng2019microwave}. A schematic of the experimental apparatus is depicted in Fig.\ref{fig:schematic} (a), in which a cutaway sectional view is projected to the left. The laser beam is focused onto the diamond with a lens 
which is mounted on a three-axis translation stage, so that the focal spot can be moved. 
Optical detection of magnetic resonance and PC readout were carried out simultaneously.

We started our measurements by characterizing an effective interrogation area for the two detection methods, by positioning the laser spot in the vicinity of the electrodes. The laser beam is focused with an estimated beam waist of $\approx6\,\mu$m and Rayleigh range $\approx200\,\mu$m. We park the spot between the electrodes in the $x$-$y$ plane and scan the beam through the sample along the $z$-axis. Figure \ref{fig:schematic} (b) depicts the recorded signals of PC and PL as the focal point crosses the diamond (note a 50\,V electric potential to the electrodes was applied in both cases to achieve comparable detection conditions). The scan in Fig.\,\ref{fig:schematic} (b) starts with the focal point outside the diamond (denoted by the amber rectangle), then proceeds to the electrode plane (front diamond surface, roughly at 0\,mm where the maximum PC occurs), continues into the diamond sample, towards its back side at 0.65\,mm, and finally, out of the diamond. The PC signal soars rapidly from nearly zero to maximum within 3-mm movement of the lens when the laser focal spot approaches the near diamond surface, and reaches a maximum at when it is focused exactly on the surface. Under this condition, the NV centers in the effective photoelectric interrogation volume experience optimal optical excitation. When the laser spot passes through the electrode plane, the PC signal falls quickly due to less intense 
photoionization which is quadratically dependent on the laser intensity, and a rapidly decaying electric field\,\cite{bourgeois2020photoelectric}. The PL signal, however, experiences a different $z$-axial profile: it slightly increases from the level of 0.6 to 0.8 when the focal spot moves from -3\,mm to -1\,mm and then falls suddenly to a level of 70\,\% to the maximum. The reasons behind involve blocking effect of the laser beam by the electrodes. Note the NV centers contributed to the PL across over the entire depth of the sample. 
PL signal gets maximized at $\sim$ 1.1\,mm where the focal spot has passed the whole diamond sample. Thereafter, it decreases to 70\,\% at 2\,mm and finally flattens out. 

From the discussion above, we conclude that the 
electrical readout provides a smaller effective interrogation volume compared to that of the fluorescing region due to the sharp electric field profile and the two-photon ionization process. Conversely, with optical detection, one collects photons from the entire illuminated volume. 
To better understand the system, we quantitatively reproduce the electric and optical signals based on a microscopic mathematical model\,\cite{emilie2022} and estimate the dimension of the effective detection volume for the two detection methods. The model potentially allows us to predict the densities of NV centers as well as optically inactive impurities [such as substitutional nitrogen atoms (P1 centers)].
 

\begin{figure*}
\centering
\includegraphics[width=1.85\columnwidth]{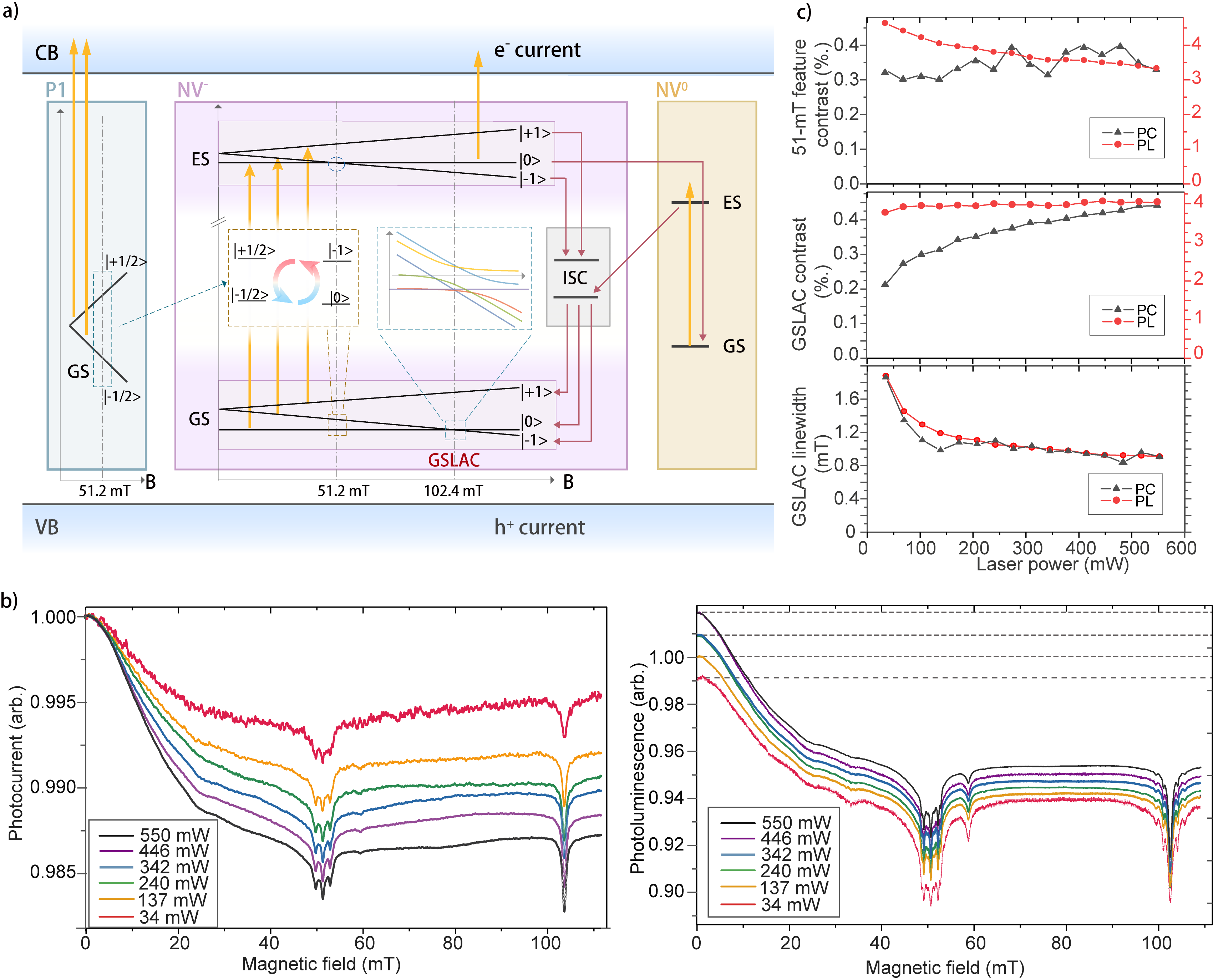}
\caption{PC detection vs magnetic field under various laser power. (a) The conventionally assumed photoionization mechanism of NV and P1 defects in diamond (an alternative mechanism has recently discussed in Ref.\,\cite{razinkovas2021photoionization}. Charge and spin initialization is carried out via 532\,nm laser excitation (amber arrow). The red arrows denote non-radiative decay of electrons to the metastable states or the neutral state (NV$^0$). ISC: Inter-system crossing. CB: conduction band. VB: valence band. e$^-$: electron. h$^+$: free hole.
(b) PC (left) and PL (right) signal as a function of magnetic field. The PC and PL signals were monitored while the magnetic field was scanned from 0 to 110 mT in 5 s. The green laser beam of various powers is modulated and the PC is synchronously detected with lock-in amplifier. The data are normalized to the signal value at zero field. The laser power is indicated in corresponding colors in the legend. For visual clarity, the traces in the right figure are manually shifted. The electrical potential applied to the electrodes is 17\,V. (c) Obtained parameters (contrast and/or full width at half maximum (FWHM)) for the features of cross relaxation and the GSLAC with photoluminescence (solid circles) and PC (solid triangles) readout as a function of the green laser power. }
\label{fig:laserpowergslacvsmagneticfield}
\end{figure*}

\textit{Photoionization within diamond}.---Figure\,\ref{fig:laserpowergslacvsmagneticfield} summarizes optical and spin transitions at the 51\,mT and the GSLAC point. Under sufficient optical illumination, defects inside semiconductor material can be ionized, thereby producing charge carries that move towards electrodes under electric field\,\cite{aslam2013photo,siyushev2013optically}. Figure\,\ref{fig:laserpowergslacvsmagneticfield} (a) illustrates the photoionization of NV centers and P1 centers.
The photoionization cycle of NV centers in diamond is spin-dependent\,\cite{razinkovas2021photoionization}. Incoming photons with the energy of 1.94\,eV or higher, can be can lead to photoionization via two photon process of the NV center\,\footnote{ If the photon energy is higher than 2.6\,eV, an electron can be promoted from the ground state ($^3A_2$) directly to the CB\,\cite{razinkovas2021photoionization}.}. In the first ionization step, the electron of the NV promoted to the excited state can decay either radiatively back to the ground state or nonradiatively into the long-lived metastable singlet state (intersystem crossing, ISC). The excited NVs can be further promoted to the conduction band by absorption of a second photon. Under an applied electric field, the electrons can freely travel in the conduction band and can finally be collected. Photoionization changes the NV charge state from NV$^-$ to NV$^0$. The recovery of the NV$^-$ center by back conversion from NV$^0$ can be induced by two-photon excitation of the NV$^0$ center and capture of an electron from the valence band\,\cite{siyushev2019photoelectrical}. 
Recently, it has been suggested that this process can also be accomplished by a single photon\,\cite{razinkovas2021photoionization,manson2018nv}. The dynamical cycling between the two charge states is enabled by optical excitation with wavelengths shorter than the zero-phonon line of NV$^0$\,\cite{siyushev2013optically}, in particular, the energies $>$ 2.6\,eV are needed for that step. 
Thus, two charge carriers are produced within one cycle: one electron during ionization and one hole during NV$^-$ recovery\,\cite{siyushev2019photoelectrical}. 
This dynamic charge circulation enables photoelectric readout of NV centers.
Photoionization inside diamond lattice can involve also other defects, e.g. P1 centers. 
The electrons from P1 centers can be promoted to the CB directly via one-photon ionization [Fig.\,\ref{fig:laserpowergslacvsmagneticfield} (a)]\,\cite{bourgeois2015photoelectric,bourgeois2020photoelectric,siyushev2013optically}. These generated charge carriers lead to a background photocurrent signal [Fig.\,\ref{fig:laserpowergslacvsmagneticfield} (a)]. 

As discussed above, the photoinduced current in ensemble experiments involves the signal produced by both NV and P1 centers. Identifying individual contributions of the two types of defects, would provide additional information useful for optimization of photoelectric readout. Towards this goal, we explored the sharp features in the magnetic-field dependence of both the PC and PL signals, corresponding to the GSLAC and other B-field regions with enhanced cross-relaxation. These are identified as being due to NV-NV\,\cite{ivady2021photoluminescence} and NV interactions with other spins such as those of P1 centers. Some of these features detected optically have already been discussed in Ref.\,\cite{pezzagna2021quantum}. In Fig.\,\ref{fig:laserpowergslacvsmagneticfield} (b) we present normalized PL and photocurrent signals as a function of the applied magnetic field under various powers of the green laser light. This figure gives an overview of the changes in the spin contrast and the resonance linewidth, of the cross-relaxation and anticrossing features. Note that the contrast is defined as $(signal_{\rm{off}}-signal_{\rm{on}})/signal_{\rm{off}}$. A remarkably sharp feature around 102.4\,mT, indicates the GSLAC which occurs when the Zeeman shift of the $|m_s=-1\rangle$ component equals to the zero-field splitting\,\cite{wickenbrock2016microwave,zheng2019microwave}. 
At around 51.2\,mT, the observed features correspond to cross-relaxation between the NV center and P1 centers \cite{wickenbrock2016microwave, Hall2016,wood2016wide,auzinsh2018hyperfine,lazda2020cross}, where these two defects can exchange energy due to resonant transition frequencies\,\cite{Hall2016}. The feature that appears near 59\,mT corresponds to cross-relaxation between on-axis (along $z$) and off-axis NVs. The sharp GSLAC and relaxation features can be used for detecting electron spin resonance (ESR) with ODMR by tuning the static external magnetic field\,\cite{wood2016wide}. In our work, we show the first electric-readout demonstration of ESR that can be used for both spin-system characterization and sensing. 

From Fig.\,\ref{fig:laserpowergslacvsmagneticfield} it is clear that the observed magnetic features show different dependence on the light power in the two detection methods. As concerns the PC readout, both the initial drop [the slope between 0 and 30\,mT, see Fig.\,\ref{fig:laserpowergslacvsmagneticfield} (b)] and the GSLAC feature [Fig.\,\ref{fig:laserpowergslacvsmagneticfield} (c, middle)] increase in magnitude and also their spectral resolution gradually increases under stronger laser excitation, while they remain nearly constant in PL readout. One reason is the photocurrent background from the non-NV defects photoionization which adds a constant photocurrent background, therefore also reducing the spin contrast. Because the NV-related features come from two-photon processes while the P1 centers responsible for the background are ionized with a single photon, the contrast of the measurement improves at higher light powers. Also, the local spin environment can be altered with the application of green light as discussed in Ref. \cite{manson2018nv}. Currently, it is not known whether such changes will affect the PL and PC readout in the same way. Furthermore, such effects could also account for the narrowing of the GSLAC features (in both methods) in the presence of light, see Fig.\,\ref{fig:laserpowergslacvsmagneticfield} (c, bottom), due to depopulation of interacting spins (P1s) in the vicinity of NV centers, since more and more impurities are photoionized. 
This might also explain why the shape and the linewidth of the cross-relaxation feature at 51 mT, see Fig.\,\ref{fig:laserpowergslacvsmagneticfield} (b), stays nearly the same if observed via PC detection. Interestingly, at the same time, the contrast reduces in the PL signal for feature at 51\,mT with increasing the laser power. We note also that the NV-NV cross-relaxation feature at 59.5\,mT and the satellite features around GSLAC, which are prominent in PL have a much lower contrast in PC. The contrast in PL detection on the NV-NV feature accounts up to 2\,\% at low laser power and decreases to 0.7\,\% at higher laser power. For PC, the contrast is too low ($<$0.1\,\%) to be quantitatively determined over the entire light-power range. For completeness, we point out the ratio of the PL contrasts of the NV-NV to NV-P1 features changes from 0.4 at low light power to 0.22 at high power. 
The difference in background signals in the overall measurement with the two methods does not fully account for the specific observation here. Understanding the reasons for these behaviors will be the subject of future work. 

\begin{figure*}
\centering
\includegraphics[width=1.8\columnwidth]{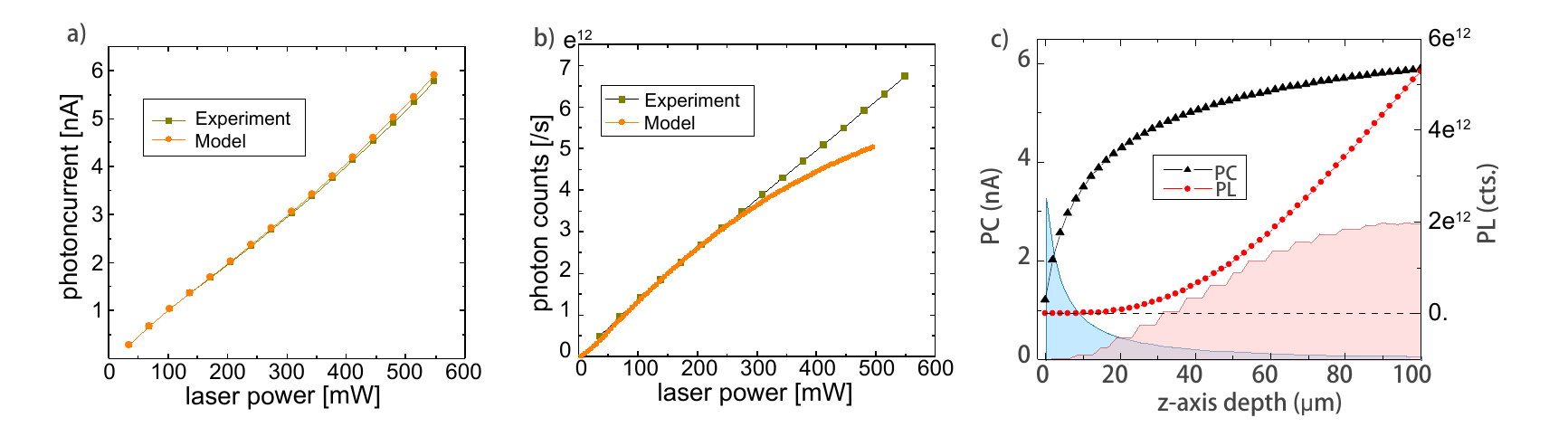}
\caption{(a). Calculated (orange) and Experimental (green) PC signal vs laser power. (b). Calculated (Orange) and Experimental (Green) PL signal vs laser power. (c) Calculated accumulated PC (black) and PL (red) signal as function of penetrating depth into diamond chip. Blue (red) shaded area indicates local contribution to the PC (PL) signal in arbitrary units.}
\label{fig:model}
\end{figure*}

To compare with the experiment, we calculate the PC and PL signal in the given experimental conditions and geometric configuration, taking into account the cross section of the Pl emission, photoionization thresholds, optical saturation, etc.\,\cite{razinkovas2021photoionization}. The rate values are taken from the published data calculations that have been verified by comparing the PL and PC traces for different laser powers\,\cite{emilie2022}. 
Figures \ref{fig:model}\,(a) and (b) show the predicted and measured laser-power dependence of the total PC and PL. Whilst the PC dependence is predicted to be saturation free, in accord with previous models\,\cite{bourgeois2015photoelectric}, PL should saturate due to the limit imposed by the lifetime of NV$^-$ excited states and shelving into metastable singlet states. The small deviation from the experiment can be accounted for by the shadowing of the diamond by the electrodes, so effectively the power density is reduced. 
Figure \ref{fig:model} (c) illustrates the simulated PC and PL signals originating from the NV centers at different depths below the surface where laser spot and electrodes are. The local contributions are indicated with blue and red shaded area (in arbitrary units). It is clear that 90\,\% PC signal comes from a region within 0-30\,$\mu$m in depth where the NV centers experience sufficient both optical excitation and electric field to drive the charge carriers towards electrodes. While photoluminescence is produced throughout the diamond slab and detected once photons get out of the diamond chip, the photocurrent only from the high E-field region beneath the contact is detected. Based on the simulations, we estimate the ratio of effective interrogation volumes of PC to PL detection to be on the order of 1:20. 


\begin{figure}
\centering
\includegraphics[width=0.8\columnwidth]{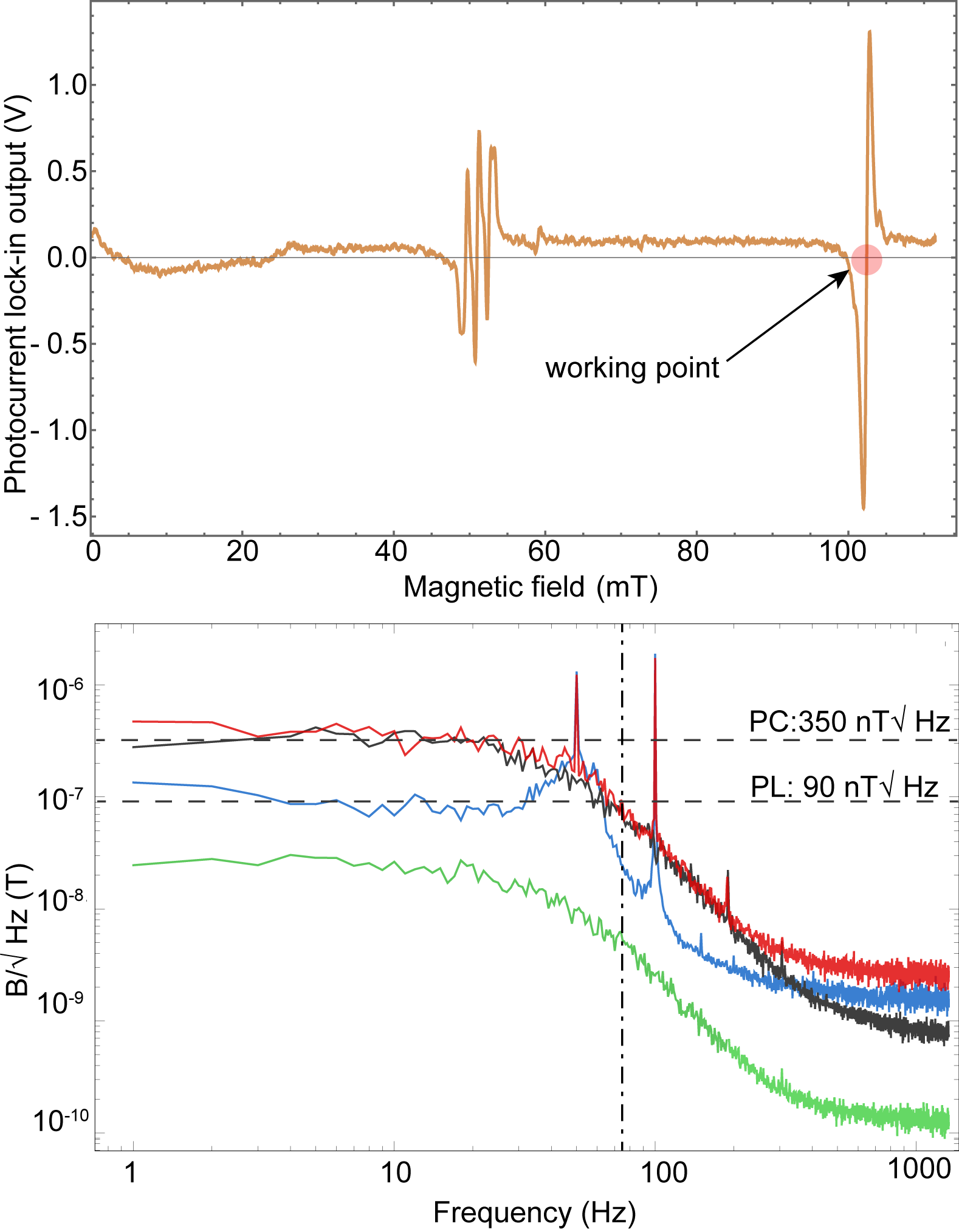}
\caption{
(a) Demodulated PC signal as a function of axial field along the $z$-axis while modulating the magnetic field. (b) Magnetic field noise spectrum. The red (blue) line indicates the noise in the magnetically sensitive configuration at a magnetic field of 102.4\,mT for PC(PL) detection, the black (green) line indicates a PC(PL) noise in the magnetically insensitive configuration (bias field $\approx$ 80\,mT). The vertical dot line indicates the bandwidth of the magnetometer. }
\label{fig:noisefloor}
\end{figure}

\textit{Electrical-readout magnetometer.} --The B-field dependent relaxation at GSLAC can be used for a realization of a microwave-free magnetometer. The principle of PL detection at the GSLAC is introduced in Ref.\,\cite{wickenbrock2016microwave}. Here we extend our approach to photocurrent readout. We apply an alternating magnetic field in addition to the bias field, and the fluorescence signal is demodulated in a properly phased LIA. The derivative signal, corresponding to the changes of the GSLAC slope with respect to the applied alternating B-field is shown in Fig.\,\ref{fig:noisefloor}\,(a) with a modulation frequency of 3.3\,kHz, and a modulation depth of $\approx$0.1\,mT which we select to be smaller than the FWHM. 
Around the GSLAC feature, the derivative signal depends approximately linearly on the magnetic field and can therefore be used for precise magnetic field measurements. The alignment and laser power were optimized to maximize the slope and the sensitivity of the magnetometer. The data near the zero-crossing are fit with a straight line to translate the LIA output signal into magnetic field (calibration). Then the background magnetic field is set to the center of the GSLAC feature (102.4\,mT) where the magnetometer is maximally sensitive to external magnetic fields. 
The sensitivity of the magnetometer can be deduced from the noise in the LIA output and the calibration. 

For noise measurements, the LIA output is recorded for 1\,s. The data are passed through a fast Fourier transform as displayed in Fig.\,\ref{fig:noisefloor}\,(b), from which we can establish the noise-floor and the sensitivity spectrum. 
For comparison, similar data are collected for both PC (blue) and PL (red) detection, at a magnetic field of the GSLAC and also around 80\,mT (black for PC and green for PL). The on-resonance noise floor is flat at around 350\,nT/$\sqrt{\text{Hz}}$ and 90\,nT/$\sqrt{\text{Hz}}$, respectively. At a field that is away from the GSLAC, the setup is insensitive to magnetic field variations and the data can be used to understand the technical noise level of the magnetometer. The off-resonance noise floor is flat at around 350\,nT/$\sqrt{\text{Hz}}$ and 30\,nT/$\sqrt{\text{Hz}}$, respectively. In the case of PC detection, the collection rate for electrons is significantly lower than the photon collection rate in the PL detection due to the mismatch of the interrogation volume, which decreases the signal to noise ratio. As mentioned, in the present geometry, the effective interrogation volume for PC detection is smaller than that for PL by a factor of 20, thus the electric-readout magnetometer would outperform a PL-based magnetometer in magnetic sensitivity per detection volume. According to the simulation, most of the incoming light penetrates to the diamond bulk and does not contribute to photocurrent generation, e.g. the large part of the photons (90\,\%) is not effectively used. This can be improved by optimizing the focusing geometry and the diamond material.

Another factor that negatively affects the PC-magnetometer sensitivity is the background signal produced by other defects, such as P1 centers\,\cite{bourgeois2015photoelectric}, which in our case reduces the contrast by a factor of 10 (see Fig.\,\ref{fig:laserpowergslacvsmagneticfield}). Incorporating a red-laser probe and two-color protocol would reduce this background and improve the magnetic sensitivity\,\cite{emilie2022}. 
Note that the NV-diamond sample used here had been used previously for the first demonstration of a microwave-free magnetometer \,\cite{wickenbrock2016microwave} that achieved 6\,nT /$\sqrt{\text{Hz}}$ noise level. As compared to \cite{wickenbrock2016microwave} a portion of the laser light is blocked by the electrodes. Improving the efficiently used photons and the spin contrast would increase the sensitivity. The main focus of this work is to demonstrate an electrical-readout NV-based magnetometer and to discuss the underlying physics and potential routes for improvement, the sensitivity will be optimized in future work.

\textit{Summary and outlook}.--
The experimental results presented in this paper demonstrate electrical detection of cross relaxation processes involving NV centers in bulk diamond. We also report calculated results based on a model\,\cite{emilie2022} that quantitatively describe and explain the experimental data. We anticipate this method, with careful and more sophisticated modelling\,\cite{emilie2022}, can be employed for quantitative measurement of photogenerated carriers, which offers a tool to determine of the concentrations of various defects, including optically inactive P1 centers, paving the way for characterizing local environment information inside diamond lattice.

We demonstrate a photoionization-based microwave-free magnetometer based on the GSLAC, which, to the best of our knowledge, is the first demonstration of electric-readout NV magnetometry that provides an alternative to conventional NV-based magnetometers and potentially overcomes some of their limitations. These results open the door to a number of intriguing future directions. First, the electrical readout is compatible with nanoscale electrode structures\,\cite{gulka2021room}, enabling advances towards integrated large-scale diamond quantum devices. It can be utilized to build scalable sensor arrays\,\cite{bourgeois2020photoelectric} and microscopes potentially superior to existing diamond quantum microscopes\,\cite{glenn2017micrometer,Lenz2021imaging} in terms of resolution. Second, it can be used to build hybrid gradiometers taking advantage of the sensing volume mismatch of PC readout and PL detection. The small sensing volume of the PC detection allows for high spatial resolution while the large sensing volume of the PL detection provides a background measurement in close proximity. 
Third, while the present studies are focused on diamond, we anticipate comparable behavior in other wide-bandgap semiconductors of technological importance.

Although we observed the magnetic resonances using red laser light (see the supplementary) that excludes the dominant current signal associated with the ionization of P1 centers, we still find a smaller GSLAC contrast\,\cite{emilie2022} for photocurrent measurements in comparison with optical detection. Further studies are necessary to reveal the sources of the remaining background photocurrent. 

\section*{Appendix}
\subsection{Experimental setup}
Figure\,\ref{fig:schematic} shows a schematic of the experimental setup. The apparatus includes a custom-built electromagnet. The electromagnet can be moved with a computer-controlled 3-D translation stage (Thorlabs PT3-Z8) and a rotation stage (Thorlabs NR360S, $x$-axis). The NV-diamond sensor is placed in the center of both the magnetic bore and a pair of integrated Helmholtz coils along the $z$-axis. The diamond can be rotated also around the $z$-axis. 
This provides all degrees of freedom for placing the diamond in the center of the magnet and aligning the NV axis parallel to the magnetic field. 

The light source is a solid-state laser emitting at a wavelength of 532\,nm (Laser Quantum Gem 532). The green laser light propagates through an acousto-optic modulator (AOM) and can be modulated by turning first-order diffraction on and off. The PL emitted by the diamond sample is collected with a 50-mm-focal-length lens and detected with a photodetector (Thorlabs APD36A).

\subsection{Diamond sample}

The sensor is constructed with a single-crystal [111]-cut $\left(2.1\times 2.3\times 0.65\right)\,\text{mm}^3$ diamond, synthesized using a high-temperature high-pressure (HPHT) method. The initial nitrogen concentration of the sample was specified as $<$200\,ppm. To produce NV centers, the sample was electron-irradiated at 14\,MeV (dose: $10^{18}\,\text{cm}^{-2}$) and then annealed at 700$^{\circ}$C for three hours. The resulting NV centers are randomly oriented along all four \{111\} crystallographic axes of the diamond. 

The electrodes for the photoelectric readout were fabricated by means of optical lithography. Prior electrode fabrication the diamond sample was cleaned in oxidizing mixture of H$_2$SO$_4$ and KNO$_3$ at $\sim$250\,$^\circ$C for approximately 30 minutes, after which it was rinsed in deionized water. The electrode structure was placed on the top surface of the diamond. Shape of the electrodes are coplanar interdigitated contacts with a gap of 5\,$\mu$m. The metal stack composition is 20\,nm titanium covered by 100\,nm aluminum for wire bonding. The electrodes are bonded to the sample holder (carrier PCB) by a 25-$\mu$m-thick aluminum wire. 

\section*{Acknowledgements}
The authors acknowledge the assistance by M. Omar and J. Shaji Rebeirro at the early stages of the project. This work was supported by the EU FET-OPEN Flagship Project ASTERIQS (action 820394), and the German Federal Ministry of Education and Research (BMBF) within the Quantumtechnologien program (FKZ 13N15064), and the Cluster of Excellence “Precision Physics, Fundamental Interactions, and Structure of Matter” (PRISMA+ EXC 2118/1) funded by the German Research Foundation (DFG) within the German Excellence Strategy (Project ID 39083149). J.H. is a PhD fellow of the Research Foundation - Flanders (FWO). P.S. acknowledges support by Baden-W{\"u}rttemberg Stiftung via Elite Program for Postdocs. M.N. acknowledges the SBO project DIAQUANT from Flemish fonds for Scientific Research No: S004018N. F.J. acknowledges the European Research Council via Synergy Grant HyperQ, the DFG via excellence cluster EXC 2154 POLiS and CRC1279, and the BMBF.

\bibliographystyle{apsrev4-2-2}
\bibliography{literature}

\section*{Supplementary}
\renewcommand{\thefigure}{S\arabic{figure}}
\setcounter{figure}{0}

\begin{figure}
\centering
\includegraphics[width=\columnwidth]{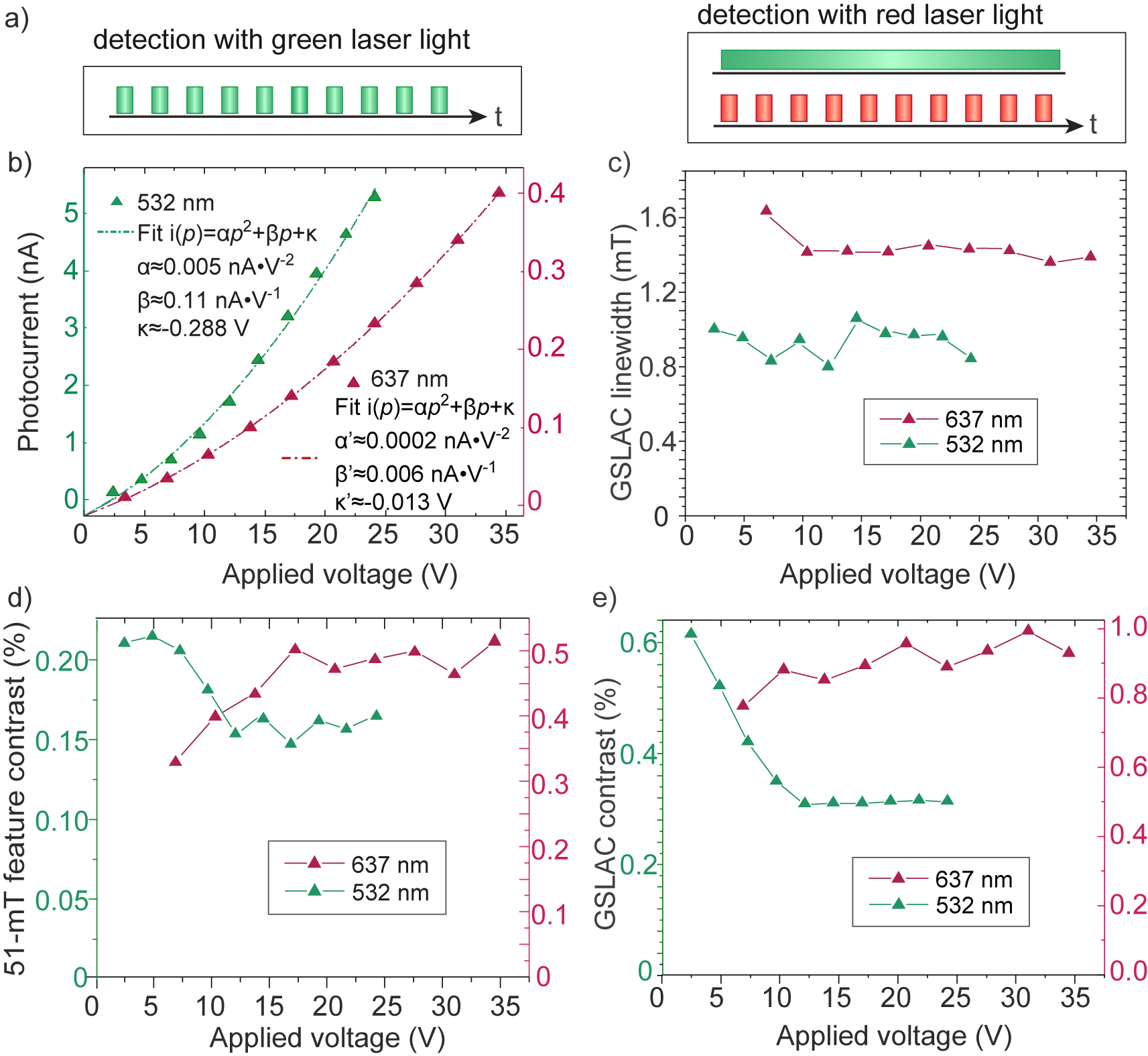}
\caption{(a). Pulsed PC readout scheme with green light (left)/red light (right). 
The 532\,nm laser remained on throughout in the red-light experiment. The green laser power is about 310\,mW and the red is 260\,mW. All light pulses were applied with a 50\% duty cycle, with a modulating frequency of 9.1\,kHz for green and 977\,Hz for red. (b). PC signal under various electrical potentials for both green- and red-light detection. Obtained parameters of PC signal, GSLAC linewidth in (c), the contrast of the 51\,mT feature and the GSLAC in (d) and (e), respectively.}
\label{fig:gslaccontrastvsbiasvolt}
\end{figure}

The electric field applied between the two electrodes is important for the readout. Here we investigate the dependence of total PC signal as well as the contrast and linewidth of the magnetic resonances on the electric potential.
We used two lasers that provide 532\,nm and 637\,nm photons, which are modulated for detection with a LIA, see Fig.\,\ref{fig:gslaccontrastvsbiasvolt} (a). The obtained parameters of the PC readout as a function of applied voltage are summarized in Fig.\,\ref{fig:gslaccontrastvsbiasvolt} (b)-(e). The acquired PC signal increases as the applied voltage increases for both excitation [Fig.\,\ref{fig:gslaccontrastvsbiasvolt} (b)].
Although the optical profile for the two lasers can not be exactly the same, 
the value of red-laser signal is less than that of the green-laser signal, 
which is consistent with the fact that in our sample. This is because the photoionization cross section for the second photoionization step for the red light is lower than for the green light leading to overall lower photocurrent\,\cite{bourgeois2020photoelectric,emilie2022}.  
Note that the red-light detection was operated with green light constantly on, for NV$^0$ recovery. 

This tendency can also be concluded from the difference in the linewidth of the GSLAC feature, shown in Fig.\,\ref{fig:gslaccontrastvsbiasvolt} (c), indicates that the green light strongly alters the local spin environment by photoionization of P1 centers, which is weaker in the case of red light excitation. 
Similar trends of the 51-mT and GSLAC features, whose contrasts decrease under green-light excitation while they moderately increase with red-light illumination shown in Fig.\,\ref{fig:gslaccontrastvsbiasvolt} (d) and (e), evidence different cycling rates of the components contributed by NV center or other impurities. 

 

\end{document}